\documentclass[aps,prd,floatfix,superscriptaddress,showpacs,showkeys]{revtex4}
\usepackage{graphicx,epsfig,epstopdf}
\usepackage{amssymb,amsmath,amsxtra,amsfonts}
\usepackage{bm}

\usepackage{longtable}
\usepackage{multirow}
\usepackage{booktabs}
\usepackage{array}
\usepackage{wrapfig}
\usepackage{color}              

\usepackage{titlesec}
\titleformat{\section}{\large\bfseries}{\thesection}{1em}{}

\newcommand{\bea}{\begin{eqnarray}}
\newcommand{\ena}{\end{eqnarray}}
\newcommand{\be}{\begin{equation}}
\newcommand{\en}{\end{equation}}
\newcommand{\nn}{\nonumber\\}
\newcommand{\ed}{\end{document}} 
\newcommand{\Tr}{\mbox{\rm{tr}}}

\begin{document}

\hfill MITP/17-059 (Mainz) 

\title{Test of the multiquark structure of $a_1(1420)$ in strong two-body decays} 

\author{Thomas~Gutsche}
\affiliation{Institut f\"ur Theoretische Physik, Universit\"at T\"ubingen,
Kepler Center for Astro and Particle Physics,
Auf der Morgenstelle 14, D-72076, T\"ubingen, Germany}

\author{Mikhail~A.~Ivanov}
\affiliation{Bogoliubov Laboratory of Theoretical Physics, 
Joint Institute for Nuclear Research, 141980 Dubna, Russia}

\author{J\"urgen~G.~K\"orner}
\affiliation{PRISMA Cluster of Excellence, Institut f\"{u}r Physik, 
Johannes Gutenberg-Universit\"{a}t,  
D-55099 Mainz, Germany}

\author{Valery~E.~Lyubovitskij}
\affiliation{Institut f\"ur Theoretische Physik, Universit\"at T\"ubingen,
Kepler Center for Astro and Particle Physics,
Auf der Morgenstelle 14, D-72076, T\"ubingen, Germany}
\affiliation{Departamento de F\'\i sica y Centro Cient\'\i fico
Tecnol\'ogico de Valpara\'\i so-CCTVal, Universidad T\'ecnica
Federico Santa Mar\'\i a, Casilla 110-V, Valpara\'\i so, Chile}
\affiliation{Department of Physics, Tomsk State University,  
634050 Tomsk, Russia} 
\affiliation{Laboratory of Particle Physics, Tomsk Polytechnic University, 
634050 Tomsk, Russia} 

\author{Kai~Xu}
\affiliation{Institut f\"ur Theoretische Physik, Universit\"at T\"ubingen,
Kepler Center for Astro and Particle Physics,
Auf der Morgenstelle 14, D-72076, T\"ubingen, Germany}
\affiliation{School of Physics and Center of Excellence in High Energy Physics 
\& Astrophysics,\\ 
Suranaree University of Technology, Nakhon Ratchasima 30000, Thailand}

\begin{abstract}

We present an analysis of strong two-body 
decays of the $a_1(1420)$ with $J^{PC} = 1^{++}$ recently 
reported by the COMPASS Collaboration at CERN. 
Following the interpretation of the COMPASS Collaboration that the $a_1$ 
is an unusual state with a four-quark $q\bar q s \bar s$ structure 
we consider two possible configurations for this state --- 
hadronic molecular and color diquark-antidiquark structures. 
We find that the dominant decay mode of the $a_1$ is the 
decay into $K$ and $K^*$. In particular, we calculate 
that the four decay modes $a_1 \to VP$ with $VP = K^{*\pm}K^\mp$, 
$K^{*0}\bar K^0$, $\bar K^{*0}K^0$ together give 
a dominant contribution to the measured total width of about 150 MeV.  
The observational mode $a_1 \to f_0(980) + \pi^0$ is significantly 
suppressed by one order of magnitude. 


\end{abstract}

\maketitle

\section{Introduction}
\label{sec:intro}

Recently the COMPASS Collaboration at CERN reported~\cite{Adolph:2015pws} 
on the observation of the new state $a_1(1420)$ with quantum numbers $J^{PC} = 1^{++}$. 
With a mass of $1414^{+15}_{-13}$ MeV and a width of $153^{+8}_{-23}$ MeV this unusual
state was interpreted by the collaboration as a tetraquark bound state
with the quark content $qs\bar q \bar s $ ($q=u$ or $d$).
The $a_1 (1420)$ was observed in the  $f_0(980) \pi^0$ decay channel, 
hinting towards the large $s\bar s$ component. 

In Ref.~\cite{Gutsche:2017oro} some of us applied holographic QCD 
to the study of the $a_1(1420)$ state. In this analysis we confirmed its
four-quark structure and predicted a mass of about
$M_{a_1} = \sqrt{2}$ GeV $\simeq 1.414$ GeV which is in good
agreement with the COMPASS result. 
The properties and possible structure interpretations of the $a_1(1420)$ state have 
also been investigated in~\cite{Wang:2014bua,Chen:2015fwa} using 
QCD sum rules and phenomenological approaches~\cite{Ketzer:2015tqa}-\cite{Wang:2015cis}.  
In particular, the results of the QCD sum rule analysis of
Ref.~\cite{Wang:2014bua} showed that the axial-vector tetraquark assignment 
to the $a_1(1420)$ is disfavored. Instead, the author argues that the 
axial-vector $a_1 (1420)$ is a mixed state of the
$a_1(1260)$ meson and a tetraquark state with the configuration
$[su]_{S=1} [\bar s \bar d]_{S=0} + [su]_{S=0} [\bar s \bar d]_{S=1}$.
Quite differently the QCD sum rule analysis performed in Ref.~\cite{Chen:2015fwa}
confirmed the existence of the $a_1(1420)$ as a tetraquark state.
Questions related to the nature of the $a_1(1420)$ meson have also been addressed  
in other papers. In particular, it has been proposed that this state 
is a consequence of rescattering effects, and in fact 
in Ref.~\cite{Ketzer:2015tqa} the $a_1(1420)$ was interpreted as
a dynamical effect due to a singularity (branch point) in the
triangle diagram formed by the processes
$a_1(1260) \to K^* \bar K$, $K^* \to K \pi$ and $K \bar K \to f_0(980)$. 
In Ref.~\cite{Basdevant:2015wma} it was shown that
a single $I=1$ spin-parity $J^{PC}=1^{++}$ $a_1$ resonance can manifest itself
as two separated mass peaks. One mass peak decays into an $S$-wave
$\rho\pi$ final state and the second one decays into a $P$-wave
$f_0(980)\pi$ system, hence the tetraquark interpretation remains unclear.
In Ref.~\cite{Liu:2015taa} it was claimed that a resonance such as the 
$a_1(1420)$ could be produced because of an
``anomalous triangle singularity'',
if the resonance is located in a specific kinematical region.
In Ref.~\cite{Aceti:2016yeb} the $a_1(1420)$ state was considered to be a
peak in the $a_1(1260) \to \pi f_0(980)$ decay mode.
Finally, in Ref.~\cite{Wang:2015cis} it was proposed to test the possible
rescattering nature of the $a_1(1420)$ in heavy meson decays.   

In this paper we test the four-quark structure interpretation of  
the $a_1(1420)$ state in the study of its strong two-body decays  
$a_1(1420) \to f_0(980) + \pi$ and $a_1(1420) \to K^{*} + K$
using the covariant confined quark model (CCQM) proposed in 
Refs.~\cite{Branz:2009cd,Dubnicka:2010kz,Dubnicka:2011mm}. 
The CCQM has been successfully 
applied to the description of the properties of the exotic 
$X(3872)$, $Z_c(3900)$, $Z(4430)$, $X(5568)$, 
$Z_b(10610)$, $Z_b'(10650)$  
states~\cite{Dubnicka:2010kz,Dubnicka:2011mm,%
Gutsche:2016cml,Goerke:2016hxf,Goerke:2017svb}. 
Our approach is based on the use 
of the compositeness condition~\cite{Weinberg:1962hj}-\cite{Efimov:1993ei} 
originally formulated in terms of hadronic~\cite{Weinberg:1962hj,Hayashi:1967hj} 
and later on in terms of quark constituents~\cite{Efimov:1993ei}.  
This procedure provides an opportunity 
to analyze both tetraquark (color diquark-antidiquark) 
and hadronic molecular configurations of the four constituent quarks 
forming the exotic state. In Refs.~\cite{Goerke:2016hxf,Goerke:2017svb} 
we have shown that the four-quark picture with molecular-type interpolating 
currents are favored for the $Z_c(3900)$, $Z_b(10610)$ and 
$Z_b'(10650)$ states. In the case of the $Z_b(10610)$ and 
$Z_b'(10650)$ states there is strong experimental confirmation~\cite{Garmash:2015rfd} 
and theoretical~\cite{Bondar:2011ev} justification that these exotic states 
are four-quark states with molecular-type interpolating currents. 
In particular, since the masses of the $Z^+_b(10610)$ and $Z_b^\prime(10650)$ 
resonances are very close to the respective $B^*\bar B$ (10604 MeV)
and  $B^*\bar B^*$ (10649 MeV) thresholds, 
in Ref.~\cite{Bondar:2011ev} it was also suggested  
that they have molecular-type binding structures.
 
We observe a similar manifestation of the hadronic molecular picture in
the case of the $a_1$ state. 
We find that the dominant decay modes of the $a_1$ are the 
$K K^\star$ channels and 
that the hadronic molecular configuration is preferred 
when compared to the color-diquark-antidiquark structure interpretation. 
In particular, we estimate 
that the four decay modes $a_1 \to VP$ with $VP = K^{*\pm}K^{\mp}$, 
$K^{*0}\bar K^0$, $\bar K^{*0}K^0$ together nearly make up the   
total width of the $a_1$ of about 150 MeV. 
The decay mode $a_1 \to f_0(980) + \pi^0$, observed by the COMPASS 
Collaboration, is significantly suppressed by one order of magnitude. 

The paper is organized as follows. In Sec.~\ref{sec:formalism}, we 
discuss the basic notions of our approach for the description of the composite 
structure of mesons and tetraquark states: the choice of interpolating
quark-antiquark and four-quark currents with the quantum numbers of the
respective states, the Lagrangians that 
describe the coupling of the bound states with their constituents, 
and the choice of parameters. 
We give model independent formulas for the matrix elements and decay rates
of the strong two-body decays of the $a_1(1420)$. 
In Sec.~\ref{sec:model-calculas} we calculate matrix elements
and decays rates in the framework of our covariant quark model.
In Sec.~\ref{sec:numerics} we discuss the numerical results obtained
in our approach and compare them with available experimental data.
Finally, in  Sec.~\ref{sec:Summary} we summarize our findings.


\section{Formalism} 
\label{sec:formalism}

According to the quantum number assignment of 
the COMPASS Collaboration~\cite{Adolph:2015pws} 
we consider the new exotic state $a_1(1420)$ as an axial-vector meson state 
with $J^{PC} = 1^{++}$, in a minimal configuration composed of a nonstrange 
and strange quark-antiquark pair. 
For the internal structure of the $a_1(1420)$  we test two possible four-quark 
configurations --- hadronic molecular (HM) and color-diquark-antidiquark (CD). 
The respective HM and CD currents for the $a_1(1420)$ state read 
\bea
{\rm HM:} \  J^\mu_{a_1} &=& \frac{1}{2} 
\biggl[ (\bar u^a \gamma^\mu s^a) (\bar s^b  \gamma^5   u^b)
      - (\bar u^a \gamma^5   s^a) (\bar s^b  \gamma^\mu u^b) 
      - (\bar d^a \gamma^\mu s^a) (\bar s^b  \gamma^5   d^b)
      + (\bar d^a \gamma^5   s^a) (\bar s^b  \gamma^\mu d^b) 
\biggr]\,, 
\label{eq:a1-curHM}\\[2mm]
{\rm CD:} \  J^\mu_{a_1} &=& \frac{\sqrt{3}}{4} \, 
\epsilon^{abd} \, \epsilon^{ced}
\biggl[   (u^a C \gamma^\mu s^b) (\bar u^c  \gamma^5 C \bar s^e)
        + (u^a C \gamma^5   s^b) (\bar u^c  \gamma^\mu C \bar s^e)
\nonumber\\
      &-& (d^a C \gamma^\mu s^b) (\bar d^c  \gamma^5 C \bar s^e)
       -  (d^a C \gamma^5   s^b) (\bar d^c  \gamma^\mu C \bar s^e)
\biggr]
\label{eq:a1-curCD}\,,  
\ena 
where $C = \gamma^0 \gamma^2$ is the charge conjugation matrix, 
and $a,b,c,d,e$ refer to color. 

The quark currents of the other hadrons relevant in the decays are specified 
as follows. We consider the $f_0(980)$ also as a four-quark state 
with $J^{PC} = 0^{++}$ applying the  two possible structures --- HM and CD: 
\bea 
{\rm HM:} \  J_{f_0} &=& \frac{1}{\sqrt{2}} 
\biggl[  (\bar u^a \gamma^5 s^a) (\bar s^b \gamma^5  u^b)
       + (\bar d^a \gamma^5 s^a) (\bar s^b \gamma^5  d^b) 
\biggr]\,, 
\label{eq:f0-curHM}\\[2mm]
{\rm CD:} \  J_{f_0} &=& \sqrt{\frac{3}{8}} \, 
\epsilon^{abd} \, \epsilon^{ced}
\biggl[  (u^a C \gamma^5 s^b) (\bar u^c  \gamma^5 C \bar s^e)
       + (d^a C \gamma^5 s^b) (\bar d^c  \gamma^5 C \bar s^e)
\biggr]\,. 
\label{eq:f0-curCD} 
\ena 
In the case of the CD currents we have added an additional factor
$\sqrt{3}/2$ resulting in
identical expressions for the mass operators 
of the four-quark states in the HM and CD versions (see below). 
In Table~\ref{tab:bb} we further specify the $J^P$ quantum numbers and the 
interpolating currents of the nonexotic quark-antiquark mesons
$\pi^0$, $K$ and $K^*$ that appear in the final decay channel. 
\begin{table}[ht]
\begin{center}
\caption{Meson states}. 
\label{tab:bb}
\def\arraystretch{1.5}
\begin{tabular}{|c|c|c|c|}
\hline
name & $J^P$ & quark current & mass (MeV) \\
\hline\hline
$\pi^0$  & $0^-$ & 
$J_{\pi^0} = 
\frac{1}{\sqrt{2}} ( \bar u i \gamma^5 u  - \bar d i \gamma^5  d)$ &
$134.977 \pm 0.0005$ \\
\hline
$K^+$  & $0^-$ & $J_{K^+} = \bar s i \gamma^5 u$
& $ 493.677\pm 0.016$ \\
\hline
$K^-$  & $0^-$ & $J_{K^-} = \bar u i \gamma^5 s$
& $ 493.677\pm 0.016$ \\
\hline
$K^0$  & $0^-$ & $J_{K^0} = \bar s i \gamma^5 d$
& $ 497.611\pm 0.013$ \\
\hline
$\bar K^0$  & $0^-$ & $J_{\bar{K}^0} = \bar d i \gamma^5 s$
& $ 497.611\pm 0.013$ \\
\hline
$K^{*+}$  & $1^-$ & $J_{K^{*+}}^\mu = \bar s \gamma^\mu u$
& $ 891.76\pm 0.25$ \\
\hline
$K^{*-}$  & $1^-$ & $J_{K^{*-}}^\mu = \bar u \gamma^\mu s$
& $ 891.76\pm 0.25$ \\
\hline
$K^{*0}$  & $1^-$ & $J_{K^{*0}}^\mu = \bar s \gamma^\mu d$
& $ 895.55\pm 0.20$ \\
\hline
$\bar K^{*0}$  & $1^-$ & $J_{\bar{K}^{*0}}^\mu = \bar d \gamma^\mu s$
& $ 895.55\pm 0.20$ \\
\hline
\end{tabular}
\end{center}
\end{table}
 
In the following we discuss the spin kinematics for the two decay modes
$1^{+} \to 0^{+} + 0^{-}$ ($a_1(1420) \to f_0(980) + \pi^0$)
and
$1^{+} \to 1^{-} + 0^{-}$ ($a_1(1420) \to K^{*\pm} + K^\mp$, 
$a_1(1420) \to K^{*0}(\bar K^{*0}) + \bar K^0(K^0)$):
 
\begin{itemize}
\item The decay $1^+\to 0^+ + 0^-$,
\end{itemize}

the 4-momenta and the Lorentz index of the polarization four-vector of 
the $a_1(1420)$ state are labeled in the transition matrix element as
\be
M=\langle 0^+ (q_1), 0^-(q_2) |\,T\,|1^+(p;\mu)\rangle \,.
\label{eq:ma1f0pi}
\en
The product of the intrinsic parities of the two final states meson is 
$(-1)$ which is opposite to the parity to the initial 
state $(+1)$. Therefore the two final states mesons must have 
odd relative orbital momenta, which in the present case must be $L=1$. 
The spins $s_1 = 0$ and $s_2 = 0$ of the two final state mesons couple to
the total spin $S=0$. Thus one has a single $(LS)$ amplitude with $(L=1,S=0)$. 
The covariant transition matrix element in terms of this amplitude is given by
\be
M\,=\,A\, q_1^\mu \varepsilon_\mu\,, 
\label{eq:Af0pi}
\en
where $\varepsilon_\mu$ is the polarization vector of the $a_1(1420)$ state. 
The amplitude $A$ is related to the helicity amplitude 
$H_{\lambda=0}$ by 
\be 
H_0 = - A \, {|\bf q_1|}\,, 
\en 
where the particles of the initial and final states are on their mass-shells
with $p^2=M^2$, $q_1^2=M_1^2$, $q_2^2=M_2^2$ and $p^\mu\varepsilon_\mu=0$;  
$|{\bf q_1}|=\lambda^{1/2}(M^2,M_1^2,M_2^2)/2M$ 
is the magnitude of the final state three-momentum 
in the rest frame of the initial particle. 
Here 
$\lambda(x,y,z) = x^2 + y^2 + z^2 - 2xy - 2xz - 2yz$ is 
the K\"allen kinematical triangle function. 

The decay rate of $1^+(p)\to 0^+(q_1) + 0^-(q_2)$ finally reads
\bea
\Gamma = 
\frac{{\bf|q_1|}^3}{24\pi M^2} \, A^2 
= \frac{\bf|q_1|}{24\pi M^2} \, |H_0|^2 \, . 
\label{eq:widtha1f0pi} 
\ena

\begin{itemize}
\item The decay $1^+\to 1^- + 0^-$,
\end{itemize}

momenta and Lorentz indices of the polarization four-vectors
in the decay are labeled according to the transition matrix element
\be
M=\langle 1^- (q_1;\delta), 0^-(q_2) |\,T\,|1^+(p;\mu)\rangle \,.
\label{eq:1m0m}
\en
Parity conservation implies even relative orbital momenta in the final state 
with $L=0,2$. 
The spins $s_1=1$ and $s_2=0$ of the two final state mesons couple 
to the total 
spin $S=1$. Thus one has the two $(LS)$ amplitudes $(L=0,S=1)$
and  $(L=2,S=1)$. The covariant expansion of the transition matrix 
is then given by
\be
M\,=\,(B\,g^{\mu \delta}+ C\,q_1^\mu q_2^\delta)\,
\varepsilon_\mu\,\varepsilon^\ast_{1\delta}\,, 
\label{eq:A3}
\en
with $p^\mu\varepsilon_\mu=0$ and $q_1^\delta\varepsilon^\ast_\delta=0$.
Alternatively one may describe the transition amplitude by the
helicity amplitudes $H_{\lambda\lambda_1}$ which can be expressed as a linear
superposition of the invariant amplitudes $B$ and $C$. One has
\be
H_{00} = -\,\frac{E_1}{M_1}\,B - \frac{M}{M_1}{|\bf q_1}|^2 \,C\,, \qquad
H_{+1+1} = H_{-1-1}=-\,B\,.
\label{eq:A5}
\en

The decay rate of $1^+(p)\to 1^-(q_1) + 0^-(q_2)$ finally reads
\bea
\Gamma &=&
\frac{\bf|q_1|}{24\pi M^2}
\Big\{ 
\Big( 3 + \frac{ {\bf|q_1|}^2}{M^2_1}\Big)\,B^2
+ (M^2+M_1^2-M_2^2)\,\frac{ {\bf|q_1|}^2}{M_1^2}\, B\,C
+ \frac{M^2}{M_1^2}{\bf|q_1|}^4\,C^2 \Big\}
\nn[2ex]
&= & \frac{\bf|q_1|}{24\pi M^2}
\Big\{ |H_{+1+1}|^2 +  |H_{-1-1}|^2 +  |H_{00}|^2 \Big\}\,.
\label{eq:width1m0m}
\ena
 
\section{Strong two-body decays of $a_1(1420)$ in the covariant quark model} 
\label{sec:model-calculas}

The nonlocal versions of the quark-antiquark currents of 
the $\pi^0$, $K$ and $K^*$ mesons written down in Table~\ref{tab:bb} 
are given by 
\bea 
\pi^0: \ J_{\pi^0}(x) &=& \int\! dy \, 
\Phi_{\pi^0}(y^2) \, J_{\pi^0}(x,y)\,, 
\label{eq:nonlocal-pi0}\\
J_{\pi^0}(x,y)&=&  \frac{1}{\sqrt{2}} 
\Big\{
  \bar u^a(x+y/2) i\gamma^5 u^a(x-y/2) 
- \bar d^a(x+y/2) i\gamma^5 d^a(x-y/2)  
\,\Big\}\,,\nonumber\\
& &\nonumber\\
K: \ J_{K}(x) &=& \int\! dy \, 
\Phi_{K}(y^2) \, J_{K}(x,y)\,, 
\label{eq:nonlocal-K}\\
J_{K^+}(x,y)     
&=& \bar s^a(x+\hat{w} y) i\gamma^5 u^a(x-\hat{w}_s y) \,, \nonumber\\
J_{K^-}(x,y)     
&=& \bar u^a(x+\hat{w}_s y)   i\gamma^5 s^a(x-\hat{w} y) \,, \nonumber\\
J_{K^0}(x,y)     
&=& \bar s^a(x+\hat{w} y) i\gamma^5 d^a(x-\hat{w}_s y) \,, \nonumber\\
J_{\bar K^0}(x,y) 
&=& \bar d^a(x+\hat{w}_s y)   i\gamma^5 s^a(x-\hat{w} y) \,, \nonumber\\
& &\nonumber\\
K^*: \ J_{K^*}^\mu(x) &=& \int\! dy \, 
\Phi_{K^*}(y^2) \, J_{K^*}^\mu(x,y)\,, 
\label{eq:nonlocal-Kstar}\\
J_{K^{*+}}^\mu(x,y) 
&=& \bar s^a(x+\hat{w} y) \gamma^\mu u^a(x-\hat{w}_s y)   \,, \nonumber\\
J_{K^{*-}}^\mu(x,y)      
&=& \bar u^a(x+\hat{w}_s y)   \gamma^\mu s^a(x-\hat{w} y) \,, \nonumber\\
J_{K^{*0}}^\mu(x,y)      
&=& \bar s^a(x+\hat{w} y) \gamma^\mu d^a(x-\hat{w}_s y) \,, \nonumber\\
J_{\bar K^{*0}}^\mu(x,y) 
&=& \bar d^a(x+\hat{w}_s y) \gamma^\mu s^a(x-\hat{w} y) \,, \nonumber
\ena 
where $\hat{w_s} = m_s/(m+m_s)$ and $\hat{w} = m/(m+m_s)$ are the fractions 
of the masses of nonstrange $m=m_u=m_d$ (we work in the isospin limit) 
and strange $m_s$ quark obeying the condition $\hat{w} + \hat{w_s} = 1$. 
The $\Phi_{\pi^0}(y^2)$, $\Phi_{K}(y^2)$, $\Phi_{K^*}(y^2)$ 
denote the set of vertex functions of the $\pi^0$, $K$, and $K^*$ states,
respectively.

The nonlocal extensions of the four-quark currents 
of the $a_1(1420)$ and $f_0(980)$ states written down in  
Eqs.~(\ref{eq:a1-curHM})-(\ref{eq:f0-curCD}) are given by 
\bea 
a_1(1420)\,, \ {\rm HM}: \ J^\mu_{a_1;HM}(x) &=& \int\! dx_1\ldots \int\! dx_4 
\delta\left(x-\sum\limits_{i=1}^4 w_i x_i\right) 
\Phi_{a_1}\Big(\sum\limits_{i<j} (x_i-x_j)^2 \Big)
J^\mu_{a_1;{\rm HM}}(x_1,\ldots,x_4)\,,
\label{eq:nonlocal-mol-cura1}\\
J^\mu_{a_1;{\rm HM}}(x_1,\ldots,x_4) &=&  \frac{1}{2} 
\Big\{
  (\bar u^a(x_3) \gamma^\mu s^a(x_1))  (\bar s^b(x_2) \gamma^5   u^b(x_4) )
- (\bar u^a(x_3) \gamma^5   s^a(x_1))  (\bar s^b(x_2) \gamma^\mu u^b(x_4) )
\nonumber\\
&-& (\bar d^a(x_3) \gamma^\mu s^a(x_1))  (\bar s^b(x_2) \gamma^5   d^b(x_4) )
  + (\bar d^a(x_3) \gamma^5   s^a(x_1))  (\bar s^b(x_2) \gamma^\mu d^b(x_4) )
\,\Big\}\,,
\nonumber
\ena 
\bea
f_0(980)\,, \ {\rm HM}: \ J_{f_0;{\rm HM}}(x) &=& \int\! dx_1\ldots \int\! dx_4 
\delta\left(x-\sum\limits_{i=1}^4 w_i x_i\right) 
\Phi_{f_0}\Big(\sum\limits_{i<j} (x_i-x_j)^2 \Big)
J_{f_0;{\rm HM}}(x_1,\ldots,x_4)\,,
\label{eq:nonlocal-mol-curf0}\\
J_{f_0;{\rm HM}}(x_1,\ldots,x_4) &=&  \frac{1}{\sqrt{2}} 
\Big\{
  (\bar u^a(x_3) \gamma^5 s^a(x_1))  (\bar s^b(x_2) \gamma^5 u^b(x_4) )
+ (\bar d^a(x_3) \gamma^5 s^a(x_1))  (\bar s^b(x_2) \gamma^5 d^b(x_4) )
\,\Big\}\,,
\nonumber
\ena 
\bea 
a_1(1420)\,, \ {\rm CD}: \ J^\mu_{a_1;{\rm CD}}(x) &=& 
\int\! dx_1\ldots \int\! dx_4 
\delta\left(x-\sum\limits_{i=1}^4 w_i x_i\right) 
\Phi_{a_1}\Big(\sum\limits_{i<j} (x_i-x_j)^2 \Big)
J^\mu_{a_1;{\rm CD}}(x_1,\ldots,x_4)\,,
\label{eq:nonlocal-cd-cura1}\\
J^\mu_{a_1;{\rm CD}}(x_1,\ldots,x_4)  
&=&  \frac{\sqrt{3}}{4} \ \epsilon^{abd} \, \epsilon^{ced} \,  
\Big\{
     (u^a(x_4) C \gamma^\mu s^b(x_1))  (\bar u^c(x_3) \gamma^5   C \bar s^e(x_2) )
\nonumber\\
 &+& (u^a(x_4) C \gamma^5   s^b(x_1))  (\bar u^c(x_3) \gamma^\mu C \bar s^e(x_2) )
\nonumber\\
 &-& (d^a(x_4) C \gamma^\mu s^b(x_1))  (\bar d^c(x_3) \gamma^5   C \bar s^e(x_2) )
\nonumber\\
 &-& (d^a(x_4) C \gamma^5   s^b(x_1))  (\bar d^c(x_3) \gamma^\mu C \bar s^e(x_2) )
\,\Big\}\,,
\nonumber
\ena 
and 
\bea 
f_0(980)\,, \ {\rm CD}: \ J^\mu_{f_0;{\rm CD}}(x) &=& 
\int\! dx_1\ldots \int\! dx_4 
\delta\left(x-\sum\limits_{i=1}^4 w_i x_i\right) 
\Phi_{f_0}\Big(\sum\limits_{i<j} (x_i-x_j)^2 \Big)
J_{f_0;{\rm CD}}(x_1,\ldots,x_4)\,,
\label{eq:nonlocal-cd-curf0}\\
J_{f_0;{\rm CD}}(x_1,\ldots,x_4) 
&=&  \sqrt{\frac{3}{8}} \ \epsilon^{abd} \, \epsilon^{ced} \,  
\Big\{
       (u^a(x_4) C \gamma^5 s^b(x_1))  (\bar u^c(x_3) \gamma^5   C \bar s^e(x_2) )
\nonumber\\
   &+& (d^a(x_4) C \gamma^5 s^b(x_1))  (\bar d^c(x_3) \gamma^5 C \bar s^e(x_2) )
\,\Big\}\,, 
\nonumber
\ena 
where $w_i = m_i/\sum\limits_{i=1}^4 m_i$ is the fraction of constituent
masses in the case of
four-quark states. Again, the vertex functions of $a_1$ and $f_0$ are denoted
by  $\Phi_H$, $H = f_0$, $a_1$.

The effective interaction Lagrangians describing the coupling of 
the $\pi^0$, $K$, $K^*$, $a_1$, and $f_0$ states 
to their constituent quarks is written in the form
\bea
{\cal L}_{ {\rm int}, \pi^0}(x) &=& 
g_{\pi^0}\,\pi^0(x) \, J_{\pi^0}(x)\,, 
\label{eq:pi-lag}\\
{\cal L}_{ {\rm int}, K}(x) &=& 
g_{K}\,K(x) \, J_{K}(x)\,, 
\label{eq:K-lag}\\
{\cal L}_{ {\rm int}, K^*}(x) &=& 
g_{K^*}\,K_\mu^*(x) \, J_{K^*}^\mu(x)\,, 
\label{eq:Kstar-lag}\\
{\cal L}_{ {\rm int}, a_1}(x) &=& 
g_{a_1}\,a_{1,\,\mu}(x)\, J^\mu_{a_1}(x)\,, 
\label{eq:a1-lag}\\     
{\cal L}_{ {\rm int}, f_0}(x) &=& g_{f_0}\, 
f_0(x) \, J_{f_0}(x) 
\label{eq:f0-lag}\,. 
\ena     
The coupling constants $g_{H}$, $H = \pi^0$, $K$, $K^*$, $a_1$, $f_0$ 
in Eqs.~(\ref{eq:pi-lag})-(\ref{eq:f0-lag}) 
are determined by the normalization condition called 
{\it the compositeness condition}~\cite{Weinberg:1962hj}-\cite{Efimov:1993ei} 
\be
\label{eq:Z=0}
Z_{H} = 1-g^2_{H}\,\widetilde\Pi_{H}^\prime(M^2_{H})=0,
\en
where $\Pi_{H}(p^2)$ is the mass operator for the $\pi^0$, $K$, and $f_0$ 
mesons and the scalar part of the $K^*$ and $a_1$ mass 
operator 
\bea
\widetilde\Pi^{\mu\nu}_{H}(p) &=& g^{\mu\nu} \widetilde\Pi_{H}(p^2)
                            + p^\mu p^\nu \widetilde\Pi^{(1)}_{H}(p^2),\nn
\widetilde\Pi_{H}(p^2) &=&
\frac13\left(g_{\mu\nu}-\frac{p_\mu p_\nu}{p^2}\right)\Pi^{\mu\nu}_{H}(p).
\label{eq:mass}
\ena

The Fourier-transforms of the $\pi^0$, $K$, $K^*$, $a_1$ and $f_0$ mass operators  
are given by 
\bea
\widetilde\Pi_{\pi^0}(p) &=& 
3 \,\int\!\!\frac{d^4k_i}{(2\pi)^4i}\,
\widetilde\Phi_{\pi^0}(-k^2) 
\,\Tr\left[\gamma_5  S(k+p/2) \gamma_5 S(k-p/2)  \right]\,, 
\label{eq:pi0-mass}\\
\widetilde\Pi_{K}(p) &=& 
3 \,\int\!\!\frac{d^4k_i}{(2\pi)^4i}\,
\widetilde\Phi_{K}(-k^2) 
\,\Tr\left[\gamma_5  S(k+p w) \gamma_5 S_s(k-p w_s)  \right]\,, 
\label{eq:K-mass}\\
 \widetilde\Pi_{K^*}^{\mu\nu}(p) &=& 
3 \,\int\!\!\frac{d^4k_i}{(2\pi)^4i}\,
\widetilde\Phi_{K^*}(-k^2) 
\,\Tr\left[\gamma^\mu  S(k+p w) \gamma^\nu S_s(k-p w_s)  \right]
\label{eq:Kstar-mass}
\ena
and for the $a_1$ as 
\bea
\widetilde\Pi_{a_1}^{\mu\nu}(p) &=& 
\frac{9}{2}\,\prod\limits_{i=1}^3\int\!\!\frac{d^4k_i}{(2\pi)^4i}\,
\widetilde\Phi_{a_1}^2\left(-\,\omega^{\,2}\right) 
\nn
&\times& \Big\{
\,\,\Tr\left[\gamma_5   S_1(\hat k_1) \gamma_5   S_3(\hat k_3) \right]
    \Tr\left[\gamma^\mu S_4(\hat k_4) \gamma^\nu S_2(\hat k_2) \right] 
\nn
&&
+\, \Tr\left[\gamma^\mu S_1(\hat k_1) \gamma^\nu S_3(\hat k_3) \right]
    \Tr\left[\gamma_5   S_4(\hat k_4) \gamma_5   S_2(\hat k_2) \right] 
\Big\}
\label{eq:a1-mass}
\ena
and for the $f_0$
\bea
\widetilde\Pi_{f_0}(p) &=& 
9 \,\prod\limits_{i=1}^3\int\!\!\frac{d^4k_i}{(2\pi)^4i}\,
  \widetilde\Phi_{f_0}^2\left(-\,\omega^{\,2}\right) 
\nn
&\times& 
\,\,\Tr\left[\gamma_5   S_1(\hat k_1) \gamma_5   S_3(\hat k_3) \right]
    \Tr\left[\gamma_5   S_4(\hat k_4) \gamma_5   S_2(\hat k_2) \right] \,. 
\label{eq:f0-mass}
\ena
In previous equations we use the notation
\bea 
& &\hat k_1=k_1-p w_1\,, \ 
   \hat k_2=k_2-p w_2\,, \ 
   \hat k_3=k_3+p w_3\,, \
   \hat k_4=k_1+k_2-k_3+p w_4\,, \nn 
& &\omega^{\,2}=\frac{1}{2}\,
\biggl(k_1^2+k_2^2+k_3^2+k_1k_2-k_1k_3-k_2k_3\biggr) 
\ena 
and $S_i(k) = 1/(m_i - \not\! k)$ is the free quark propagator with 
constituent mass $m_i$. In particular, 
$S(k)$ and $S_s(k)$ are the nonstrange and strange quark propagators. 
$\widetilde\Phi_{H}(-k^2) = \exp(k^2/\Lambda_H^2)$ is the Fourier-transform 
of the vertex function of the hadron $H$, where $\Lambda_H$ is the hadronic 
size parameter. 
Note that the mass operators of the four-quark states $a_1$ and $f_0$ 
are formally identical for both the HM and CD currents. 
  
Next we list the matrix elements of the two-body decays 
of the $a_1(1420)$ state. In case of the transition
$a_1(1420) \to f_0(980) + \pi^0$ there are four cases 
depending on the structure assumptions: 
(1) $a_1$ and $f_0$ are the HM states (HM $\to$ HM transition), 
(2) $a_1$ and $f_0$ are the CD states (CD $\to$ CD transition), 
(3) $a_1$ is the HM and $f_0$ is the CD state (HM $\to$ CD transition), 
(4) $a_1$ is the CD and $f_0$ is the HM state (CD $\to$ HM transition). 
The matrix elements for the HM $\to$ HM and CD $\to$ CD transitions 
are apart from the spatial correlations contained in the vertex function 
identical and are given by 
\bea
&&
M^{\mu; (1)}\left( a_1(p,\mu) \to f_0(q_1)+\pi^0(q_2)\right)
= 9\,g_{a_1} \, g_{f_0} \, g_{\pi^0} \nonumber\\
&\times&
\int\!\!\frac{d^4k_1}{(2\pi)^4i}\,
\int\!\!\frac{d^4k_2}{(2\pi)^4i}\,
\int\!\!\frac{d^4k_3}{(2\pi)^4i}\,
\widetilde\Phi_{a_1}\left(-\omega_{a_1}^{\,2}\right)
\widetilde\Phi_{f_0}\left(-\omega_{f_0}^{\,2}\right)
\widetilde\Phi_{\pi^0}\left(-\omega_{\pi^0}^{\,2}\right)
\nn
&\times& 
\,\,\Tr\left[ \gamma^5   S_3(\hat{k}_3') \gamma^5 S_3(\hat{k}_3) 
              \gamma^\mu S_1(\hat{k}_1)\gamma_5 \right] \ 
    \Tr\left[ \gamma^5   S_4(\hat{k}_4) \gamma^5 S_2(\hat{k}_2) \right]  
\nn[2ex]
&=& A_{a_1f_0\pi}^{(1)}\, q_1^\mu \,. 
\label{eq:decay-diag-a1f0pi}
\ena 
The matrix elements for the HM $\to$ CD and CD $\to$ HM transitions are 
also the same and are given by 
\bea
&&
M^{\mu; (2)}\left( a_1(p,\mu) \to f_0(q_1)+\pi^0(q_2)\right)
=  - 3 \sqrt{3} \,g_{a_1} \, g_{f_0} \, g_{\pi^0} \nonumber\\
&\times&
\int\!\!\frac{d^4k_1}{(2\pi)^4i}\,
\int\!\!\frac{d^4k_2}{(2\pi)^4i}\,
\int\!\!\frac{d^4k_3}{(2\pi)^4i}\,
\widetilde\Phi_{a_1}\left(-\omega_{a_1}^{\,2}\right)
\widetilde\Phi_{f_0}\left(-\omega_{f_0}^{\,2}\right)
\widetilde\Phi_{\pi^0}\left(-\omega_{\pi^0}^{\,2}\right)
\nn
&\times& 
\,\,\biggl\{\Tr\left[ \gamma^5   S_3(\hat{k}_3') \gamma^5 S_3(\hat{k}_3) 
                      \gamma^\mu S_1(\hat{k}_1)  \gamma_5 S_4(\hat{k}_4) 
                      \gamma^5   S_2(\hat{k}_2)  \right]  \nonumber\\
&+& 
 \,\,\Tr\left[ \gamma^5   S_3(\hat{k}_3') \gamma^5 S_3(\hat{k}_3) 
                      \gamma^5   S_1(\hat{k}_1)  \gamma_5 S_4(\hat{k}_4) 
                      \gamma^\mu S_2(\hat{k}_2)  \right]  
\biggr\}
\nn[2ex]
&=& A_{a_1f_0\pi}^{(2)}\, q_1^\mu \,. 
\label{eq:decay-nondiag-a1f0pi}
\ena
In both Eqs.~(\ref{eq:decay-diag-a1f0pi}) and (\ref{eq:decay-nondiag-a1f0pi}) 
we use the same notation as in the expression for the mass operators with
the additional quantities  
\bea  
\hat{k}_3' &=& \hat{k}_3 - q_2\,,   
\ena 
and   
\bea
\omega_{a_1}^{\,2} &=& \frac{1}{2} \biggl(k_1^2 + k_2^2 + k_3^2 
+ k_1k_2 - k_1k_3 - k_2k_3\biggr)\,, \nonumber\\ 
\omega_{f_0}^{\,2} &=& \frac{1}{2} \biggl(k_1^2 + k_2^2 + {k_3^\prime}^2 
+ k_1k_2 - k_1k_3' - k_2k_3' \biggr)\,, \\ 
\omega_{\pi^0}^{\,2} &=& \biggl(k_3 + p w_3 - \frac{q_2}{2}\biggr)^2 
\,. \nonumber
\ena  
The matrix elements for the $a_1(1420) \to K^* + K$ transitions obey the
following conditions due to charge conjugation symmetry 
\bea 
& &  M^{\mu\delta}\left(a_1(p,\mu) \to K^{*+}(q_1,\delta)+K^-(q_2)\right) = 
   - M^{\mu\delta}\left(a_1(p,\mu) \to K^{*-}(q_1,\delta)+K^+(q_2)\right) 
\,,\nonumber\\
& &  M^{\mu\delta}\left(a_1(p,\mu) \to K^{*0}(q_1,\delta)+\bar K^0(q_2)\right) = 
   - M^{\mu\delta}\left(a_1(p,\mu) \to \bar K^{*0}(q_1,\delta)+K^0(q_2)\right) \,.
\ena 
The matrix elements of the decays 
$a_1 \to K^{*+} + K^-$ and 
$a_1 \to K^{*0} + \bar{K}^0$ read 
\bea
{\rm HM}: \ &&
M^{\mu\delta}
\left(a_1(p,\mu) \to K^{*+}(q_1,\delta)+K^-(q_2)\right) 
=\frac{9}{2} \,g_{a_1} \, g_{K^{*+}} g_{K^-}
\nn
&\times&
\int\!\!\frac{d^4k_1}{(2\pi)^4i}\,\int\!\!\frac{d^4k_2}{(2\pi)^4i}\,
\widetilde\Phi_{a_1}\left(-\,\delta_{a_1}^{\,2}\right)
\widetilde\Phi_{K^*}\left(-\,\delta_{K^*}^{\,2}\right)
\widetilde\Phi_{K}\left(-\,\delta_{K}^{\,2}\right)
\nn
&\times& 
\,\,\Tr\left[\gamma^\mu S_1(k_1) \gamma^\delta S_3(k_1+q_1) \right] \ 
    \Tr\left[\gamma_5   S_2(k_2) \gamma_5      S_4(k_2+q_2) \right] 
\nn
&=& B_{a_1  K^{*+} K^-}\,g^{\mu\delta} 
  + C_{a_1 K^{*+} K^-}\,q_1^\mu q_2^\delta\,,
\label{eq:decay-HM-a1KstarplusKminus}
\\
{\rm HM}: \ &&
M^{\mu\delta}
\left(a_1(p,\mu) \to K^{*0}(q_1,\delta)+\bar K^0(q_2)\right) 
= - \frac{9}{2} \,g_{a_1} \, g_{K^{*0}} g_{K^0}
\nn
&\times&
\int\!\!\frac{d^4k_1}{(2\pi)^4i}\,\int\!\!\frac{d^4k_2}{(2\pi)^4i}\,
\widetilde\Phi_{a_1}\left(-\,\delta_{a_1}^{\,2}\right)
\widetilde\Phi_{K^*}\left(-\,\delta_{K^*}^{\,2}\right)
\widetilde\Phi_{K}\left(-\,\delta_{K}^{\,2}\right)
\nn
&\times& 
\,\,\Tr\left[\gamma^\mu S_1(k_1) \gamma^\delta S_3(k_1+q_1) \right] \ 
    \Tr\left[\gamma_5   S_2(k_2) \gamma_5      S_4(k_2+q_2) \right] 
\nn
&=& B_{a_1  K^{* 0} \bar K^0}\,g^{\mu\delta} 
 +  C_{a_1  K^{* 0} \bar K^0}\,q_1^\mu q_2^\delta 
\label{eq:decay-HM-a1Kstar0barK0}
\\
{\rm CD}: \ &&
M^{\mu\delta}
\left(a_1(p,\mu) \to K^{*+}(q_1,\delta)+K^-(q_2)\right) 
= - \frac{3 \sqrt{3}}{2} \,g_{a_1} \, g_{K^{*+}} g_{K^-}
\nn 
&\times&
\int\!\!\frac{d^4k_1}{(2\pi)^4i}\,\int\!\!\frac{d^4k_2}{(2\pi)^4i}\,
\widetilde\Phi_{a_1}\left(-\,\delta_{a_1}^{\,2}\right)
\widetilde\Phi_{K^*}\left(-\,\delta_{K^*}^{\,2}\right)
\widetilde\Phi_{K}\left(-\,\delta_{K}^{\,2}\right)
\nn
&\times& 
\,\,\biggl\{\Tr\left[\gamma^\mu S_1(k_1) \gamma^\delta S_3(k_1+q_1) 
             \gamma^5   S_2(k_2) \gamma^5  S_4(k_2+q_2) \right] 
\nn
&+& 
\,\,\Tr\left[\gamma^5 S_1(k_1) \gamma^\delta S_3(k_1+q_1) 
         \gamma^\mu   S_2(k_2) \gamma^5      S_4(k_2+q_2) \right] \biggr\}
\nn
&=& B_{a_1  K^{*+} K^-}\,g^{\mu\delta} 
 +  C_{a_1 K^{*+} K^-}\,q_1^\mu q_2^\delta\,,
\label{eq:decay-CD-a1KstarplusKminus}
\\
{\rm CD}: \ &&
M^{\mu\delta}
\left(a_1(p,\mu) \to K^{*0}(q_1,\delta)+\bar K^0(q_2)\right) 
= \frac{3 \sqrt{3}}{2} \,g_{a_1} \, g_{K^{*0}} g_{K^0}
\nn
&\times&
\int\!\!\frac{d^4k_1}{(2\pi)^4i}\,\int\!\!\frac{d^4k_2}{(2\pi)^4i}\,
\widetilde\Phi_{a_1}\left(-\,\delta_{a_1}^{\,2}\right)
\widetilde\Phi_{K^*}\left(-\,\delta_{K^*}^{\,2}\right)
\widetilde\Phi_{K}\left(-\,\delta_{K}^{\,2}\right)
\nn
&\times& 
\,\,\biggl\{\Tr\left[\gamma^\mu S_1(k_1) \gamma^\delta S_3(k_1+q_1) 
                     \gamma^5   S_2(k_2) \gamma^5      S_4(k_2+q_2) \right] 
\nn
&+& 
\,\,\Tr\left[\gamma^5 S_1(k_1) \gamma^\delta S_3(k_1+q_1) 
         \gamma^\mu   S_2(k_2) \gamma^5      S_4(k_2+q_4) \right] \biggr\}
\nn
&=& B_{a_1  K^{* 0} \bar K^0}\,g^{\mu\delta} +  
    C_{a_1  K^{* 0} \bar K^0}\,q_1^\mu q_2^\delta\,, 
\label{eq:decay-CD-a1Kstar0barK0}
\ena
where 
\bea 
\delta_{a_1}^{\,2} &=& 
\frac{1}{8} \biggl(k_1 - k_2 + q_1 (w_1 - w_2) 
                             + q_2 (w_1 - w_2)\biggr)^2
\nonumber\\
&+& 
\frac{1}{8} \biggl(k_1 - k_2 + q_1 ( 1 + w_4 - w_3 ) 
                             + q_2 (-1 + w_4 - w_3 )\biggr)^2
\nonumber\\
&+& 
\frac{1}{4} \biggl(k_1 + k_2 + q_1 (w_1 + w_2) 
                             + q_2 (w_1 + w_2)\biggr)^2\,, 
\nonumber\\
\delta_{K^*}^{\,2} &=& (k_1+q_1 \hat{w}_1)^2\,, \quad 
          \hat{w}_1 = \frac{m_1}{m_1+m_3}\,, 
\nonumber\\
\delta_{K}^{\,2}   &=& (k_2+q_2 \hat{w}_2)^2\,, \quad 
          \hat{w}_2 = \frac{m_2}{m_2+m_4}\,. 
\ena 
The quark masses are specified as $m_1=m_2=m_s$,  $m_3=m_4=m_d=m_u$. 

\section{Numerical results}
\label{sec:numerics}

First of all, most of the adjustable parameters
of our model (constituent quark masses $m_u = m_d$ and $m_s$, 
infrared cutoff $\lambda$ and size
parameters $\Lambda_\pi$, $\Lambda_K$, and $\Lambda_{K^*}$ 
of the $\pi$, $K$, and $K^*$ mesons) have been fixed in previous 
studies (see, e.g., Refs.~\cite{Gutsche:2015mxa,Issadykov:2015iba}) 
by a global fit to a multitude of data~\cite{Olive:2016xmw}: 
\bea 
& &
m_u = m_d = 241.29 \ {\rm MeV}\,, \quad  
m_s = 428.20 \ {\rm MeV}\,, \quad 
\lambda = 181 \ {\rm MeV}\,, \nonumber\\
& &
\Lambda_\pi   =  870.77 \ {\rm MeV}\,, \quad 
\Lambda_K     = 1014.20 \ {\rm MeV}\,, \quad 
\Lambda_{K^*} =  804.82 \ {\rm MeV}\,. 
\ena 
For the  hadronic masses in our calculations 
we use the central data values~\cite{Olive:2016xmw} with: 
\bea
& &
M_{\pi^0} = 134.9766 \ {\rm MeV}\,, \quad  
M_{f_0(980)}  =  990 \ {\rm MeV}\,, \quad  
M_{a_1(1420)} = 1414 \ {\rm MeV}\,, \nonumber\\
& &
M_{K^\pm} = 493.677  \ {\rm MeV}\,, \quad  
M_{K^0} = M_{\bar K^0} = 497.611  \ {\rm MeV}
\,, \nonumber\\ 
& &
M_{K^{* \pm}} = 891.76 \ {\rm MeV}\,, \quad  
M_{K^{* 0}} = M_{\bar K^{* 0}} = 895.55 \ {\rm MeV}
\,. 
\ena 
The only two new parameters are the size parameters of the $a_1(1420)$ 
and $f_0(980)$ states $\Lambda_{a_1}$ and $\Lambda_{f_0}$. 
According to our experience in the description of light 
hadronic systems they should be of the order of 1 GeV. 

First, we analyze the decay mode $a_1(1420) \to f_0(980) + \pi^0$.  
Since this decay proceeds in a $p$-wave in the final state, this mode 
is suppressed by a factor of ${\bf|q_1|}^{2L+1}$ with $L=1$ 
where ${\bf q_1}$ is the three-momentum in the decay channel. 
We remind the reader that the corresponding decay width is 
given by $\Gamma = {\bf|q_1|}^3/(24\pi M^2) \, A^2$,  
where ${\bf|q_1|}^3/(24 \pi M^2) \simeq 0.26$ MeV. 
To obtain a total width of $\Gamma_{a_1} \sim 100$ MeV the 
dimensionless amplitude $A$ (which in fact is an effective 
$a_1f_0\pi^0$ coupling) should be relatively large, 
i.e. of the order of 20. 
Varying the parameters $\Lambda_{a_1}$ and $\Lambda_{f_0}$ 
from 0.5 to 1.6 GeV we find that the decay width of $a_1 \to f_0 + \pi^0$ 
changes in the cases where both $a_1$ and $f_0$ are hadronic molecular 
or color diquark-antidiquark four quark states as 
$\Gamma(a_1 \to f_0 + \pi^0) = 5.1^{+4.7}_{-4.9}$~MeV. 
The central value corresponds to the case 
$\Lambda_{a_1} = \Lambda_{f_0} = 1$ GeV, 
while the maximal value results in 
$\Lambda_{a_1} = \Lambda_{f_0} = 0.5$ GeV, and 
for the minimal one in 
$\Lambda_{a_1} = 0.5$ GeV and $\Lambda_{f_0} = 1.6$ GeV. 
In the case when one of these two states has a hadronic molecular 
structure and the other --- color diquark-antidiquark --- the result 
for the decay width reads 
$\Gamma(a_1 \to f_0 + \pi^0) = 1.2^{+1.6}_{-1.1}$ MeV. 
For transparency, in Fig.~1 we present three-dimensional plots for the decays 
rates 
$\Gamma(a_1 \to f_0 + \pi^0)$ as function of the dimensional parameters 
$\Lambda_{a_1}$ and $\Lambda_{f_0}$ running from 0.5 to 1.6 GeV 
for HM (CD) $\to$ HM (CD) transition (upper plot) and 
HM (CD) $\to$ CD (HM) transition (lower plot). 

Next we look at the results for the $a_1 \to K^* + K$ decays. 
Even without doing an explicit calculation it is clear that the rate for the 
color diquark-antidiquark configuration
should be suppressed in comparison with 
the hadronic molecular configuration. This expectation is supported by our 
experience in the study of decay properties of the heavy 
four-quark states $Z_c$ and $Z_b$~\cite{Goerke:2016hxf,Goerke:2017svb}. 
Our calculations for the $a_1 \to K^* + K$ decay rate also confirm this 
expectation. The rates in the color diquark-antidiquark scenario 
are suppressed by one order of magnitude in comparison with the 
hadronic molecular scenario. We found that results for the 
partial $a_1 \to \bar K^* K$ decay rates are very sensitive to a choice of the 
size parameter $\Lambda_{a_1}$ and decrease when the size parameter increases. 
Second, the results for the HM interpretation for $a_1(1420)$ are 
overestimated in the region of $\Lambda_{a_1}$ from 0.5 to 1.5 GeV. 
For convenience, we show our numerical results 
in the two regions --- from 0.5 to 1.5 GeV and 1.5 to 1.6 GeV. 
The results for some specific values of the size parameter 
$\Lambda_{a_1} = 0.5 - 1.5$~GeV and 
$\Lambda_{a_1} = 1.5 - 1.6$~GeV are shown 
in Tables~\ref{tab:theory1} and~\ref{tab:theory2}, respectively.  
We display results for the hadronic molecular 
scenario of the $a_1$ state, while the results in case 
of the color diquark-antidiquark configuration are shown in brackets. 
We present the results for four partial decay modes and for the sum of 
these four modes (last column).  
For a size parameter of 
$\Lambda_{a_1} = 1.56-1.58$ GeV we are able to describe the data 
for the total width of the $a_1(1420)$ assuming that the four modes 
$a_1 \to VP$ with $VP = K^{*\pm}K^\mp$, 
$K^{*0}\bar K^0$, $\bar K^{*0}K^0$ make up the dominant
part of the total width of the $a_1$, 
i.e. about 150 MeV (as measured by the 
COMPASS Collaboration~\cite{Adolph:2015pws}).   
If the same value of $\Lambda_{a_1}$ is used in the
evaluation 
of the decays $a_1(1420)$ into $f_0(980) \pi^0$ and $K^* K$, 
the $a_1(1420) \to f_0(980) + \pi^0$
decay width is suppressed in comparison with the 
$a_1(1420) \to K^* + K$ mode by one order of magnitude. 

\begin{figure}
\begin{center}
\epsfig{figure=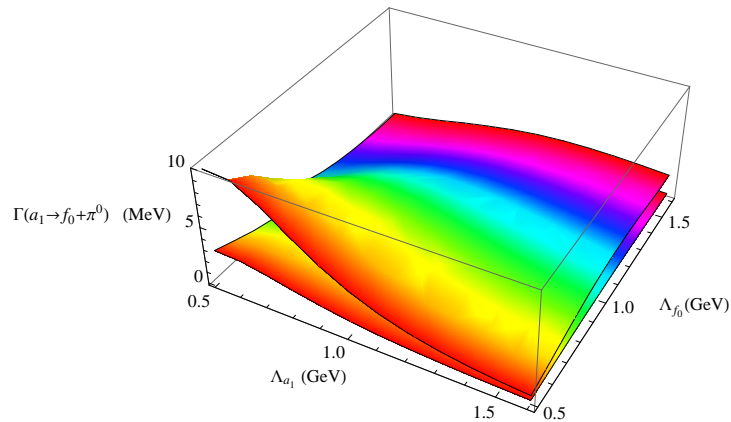,scale=.75}
\end{center}
\noindent
\caption{$\Gamma(a_1 \to f_0 + \pi^0)$ 
as function of the dimensional parameters
$\Lambda_{a_1}$ and $\Lambda_{f_0}$ running from 0.5 to 1.6 GeV
for HM (CD) $\to$ HM (CD) transition (upper plot) and
HM (CD) $\to$ CD (HM) transition (lower plot).} 
\end{figure}

\begin{table}[htp]
\centering
\caption{$a_1(1420) \to K^* + K$ decay widths (in MeV) 
for $\Lambda_{a_1} = 0.5-1.5$ GeV. 
Results for the HM structure of the $a_1$ are given first 
while the ones for the CD structure assumption are 
quoted in brackets. } 
\medskip
\label{tab:theory1}
\def\arraystretch{1}
  \begin{tabular}{|l|c|c|c|}  
\hline 
$\Lambda_{a_1}$ (GeV) &  $\Gamma(a_1 \to K^{*\pm} + K^{\mp})$ 
                &  $\Gamma(a_1 \to K^{*0}(\bar{K}^*) + \bar K^0(K^0))$ 
                &  $\Gamma_{\rm sum}(a_1 \to K^* + K)$ \\
\hline\hline  
0.50 & 1004.86 (143.57) & 880.54 (125.14) & 3770.80 (537.42) \\
0.75 &  539.84  (69.53) & 468.02  (60.04) & 2015.72 (259.14) \\
1.00 &  242.90  (29.80) & 209.67  (25.64) &  905.14 (110.88) \\
1.25 &  107.25  (12.82) &  92.37  (11.01) &  399.24  (47.66) \\
1.50 &   49.30   (5.81) &  42.40   (4.98) &  183.40  (21.58) \\
\hline 
\end{tabular}

\centering
\caption{$a_1(1420) \to K^* + K$ decay widths (in MeV) 
for $\Lambda_{a_1} = 1.5-1.6$ GeV. 
Results for the HM structure of the $a_1$ are given first 
while the ones for the CD structure assumption are 
quoted in brackets.} 
\medskip
\label{tab:theory2}
\def\arraystretch{1}
  \begin{tabular}{|l|c|c|c|}  
\hline 
$\Lambda_{a_1}$ (GeV) &  $\Gamma(a_1 \to K^{*\pm} + K^{\mp})$ 
                &  $\Gamma(a_1 \to K^{*0}(\bar{K}^*) + \bar K^0(K^0))$ 
                &  $\Gamma_{\rm sum}(a_1 \to K^* + K)$ \\
\hline\hline  
1.50 & 49.30 (5.81) & 42.40 (4.98) & 183.40 (21.58) \\
1.51 & 47.84 (5.63) & 41.15 (4.83) & 177.98 (20.92) \\
1.52 & 46.43 (5.44) & 39.94 (4.67) & 172.74 (20.22) \\
1.53 & 45.07 (5.30) & 38.77 (4.55) & 167.68 (19.70) \\
1.54 & 43.76 (5.14) & 37.63 (4.41) & 162.78 (19.10) \\
1.55 & 42.48 (4.99) & 36.54 (4.28) & 158.04 (18.54) \\
1.56 & 41.25 (4.85) & 35.47 (4.15) & 153.44 (18.00) \\
1.57 & 40.05 (4.70) & 34.45 (4.03) & 149.00 (17.46) \\
1.58 & 38.90 (4.57) & 33.45 (3.91) & 144.70 (16.96) \\
1.59 & 37.78 (4.43) & 32.49 (3.80) & 140.54 (16.46) \\
1.60 & 36.70 (4.30) & 31.55 (3.69) & 136.50 (15.98) \\
\hline 
\end{tabular}
\end{table}

Finally, we would like to discuss how the obtained results may be used
for interpretation of the observed $a_1(1420)$ meson which is supposed
to be $qs\bar q \bar s$ state.  As was pointed out in the 
review~\cite{Nielsen:2009uh} there is no one to one correspondence
between the current and the state since the CD current can be rewritten
in terms of a sum over molecular type currents through the Fierz 
transformation. However, they have shown that the molecular components appear 
with the color and Dirac suppression factors. This means that if the physical
state is a molecular state, it would be best to choose the HM current,
and vice versa, for a tetraquark state it would be better to choose 
a tetraquark current. In some sense  we follow this strategy. We are using both
the HM and CD currents to evaluate the observed decay widths. Then we simply
choose the current which is more suitable from the point of view
of the experimental data.

\section{Summary}
\label{sec:Summary}

We have tested the possible four-quark configuration of the $a_1(1420)$ state 
by studying its strong two-body decay rates for the modes
$a_1(1420) \to f_0(980) + \pi^0$ and 
$a_1(1420) \to K^* + K$. For both four-quark states $a_1(1420)$
and $f_0(980)$  
we have considered two possible structure scenarios --- 
the hadronic molecular and the color diquark-antidiquark current structure. 
We have found that the $a_1(1420) \to f_0(980) + \pi^0$
decay width is significantly suppressed in comparison to the 
$a_1(1420) \to K^* + K$ mode by one order of magnitude. 
Studying the decay $a_1(1420) \to K^* + K$ and using data of the COMPASS 
Collaboration~\cite{Adolph:2015pws} we have shown that the hadronic molecular 
configuration is preferred when compared to the compact colored diquark-antidiquark state.
In particular, for values of the size parameter
$\Lambda_{a_1} = 1.56-1.58$ GeV 
we are able to describe the available data for the total decay width of the
$a_1$ state in terms of 
the decay modes $a_1(1420) \to K^* + K$ alone: 
$\Gamma_{\rm sum}(a_1 \to K^* + K) = 144.70 - 153.44$ MeV. 

\begin{acknowledgments}

This work was supported
by the German Bundesministerium f\"ur Bildung und Forschung (BMBF)
under Project 05P2015 - ALICE at High Rate (BMBF-FSP 202):
``Jet- and fragmentation processes at ALICE and the parton structure 
of nuclei and structure of heavy hadrons'', 
by CONICYT (Chile) PIA/Basal FB0821, 
by Tomsk State University Competitiveness 
Improvement Program and the Russian Federation program ``Nauka'' 
(Contract No. 0.1764.GZB.2017), 
by Tomsk Polytechnic University Competitiveness Enhancement Program 
(Grant No. VIU-FTI-72/2017). 
M.A.I.\ acknowledges the support from  PRISMA cluster of excellence 
(Mainz Uni.). M.A.I. and J.G.K. thank the Heisenberg-Landau Grant for
partial support.  
K.X. was supported by the SUT-PhD/13/2554 Scholarship of
Suranaree University of Technology and by the Higher Education Research
Promotion and National Research University Project of Thailand, Office of
the Higher Education Commission.

\end{acknowledgments}

\end{document}